\documentclass[onecolumn,unpublished,a4paper]{quantumarticle}
\usepackage{amsmath}
\usepackage{amssymb}
\usepackage{amsthm}
\usepackage{amsfonts}
\usepackage[caption=false]{subfig}
\usepackage[colorlinks]{hyperref}
\usepackage[all]{hypcap}
\usepackage{tikz}
\usepackage{relsize}
\usepackage{color,soul}
\usepackage[utf8]{inputenc}
\usepackage{capt-of}
\usepackage{mathtools}
\usepackage[numbers,sort&compress]{natbib}
\usepackage{float}
\usepackage[section]{placeins}
\usepackage{listings}
\usepackage[T1]{fontenc}  
\usetikzlibrary{decorations.pathreplacing}
\usepackage{braket}

\DeclareFixedFont{\ttb}{T1}{txtt}{bx}{n}{4}
\DeclareFixedFont{\ttm}{T1}{txtt}{m}{n}{4}
\definecolor{deepblue}{rgb}{0,0,0.5}
\definecolor{deepred}{rgb}{0.6,0,0}
\definecolor{deepgreen}{rgb}{0,0.5,0}
\newcommand\cppstyle{\lstset{
language=C++,
basicstyle=\ttm,
otherkeywords={uint8_t, __m256i, size_t, ASSERT_TRUE, EXPECT_TRUE, TEST, BENCHMARK},
keywordstyle=\ttb\color{deepblue},
emphstyle=\ttb\color{deepblue},
stringstyle=\color{deepgreen},
commentstyle=\fontfamily{txtt}\selectfont\color{gray},
showstringspaces=false,
literate={*}{{\char42}}1
         {-}{{\char45}}1
}}
\lstnewenvironment{cpp}[1][]
{\cppstyle\lstset{#1}}{}

\newcommand\pythonstyle{\lstset{
language=python,
basicstyle=\ttm,
morekeywords={assert,as,echo},
keywordstyle=\ttb\color{deepblue},
emphstyle=\ttb\color{deepblue},
stringstyle=\color{deepgreen},
commentstyle=\fontfamily{txtt}\selectfont\color{gray},
showstringspaces=false,
literate={*}{{\char42}}1
         {-}{{\char45}}1
}}
\lstnewenvironment{python}[1][]
{\pythonstyle\lstset{#1}}{}

\lstdefinestyle{qsharp}{
    language=go,
    basicstyle=\fontsize{8}{8}\selectfont\ttfamily,
    upquote=true,
    stepnumber=1,
    numbersep=8pt,
    showstringspaces=false,
    breaklines=true,
    frame=single,
    aboveskip=1.5em,
    belowskip=1.5em,
    commentstyle=\color{gray},
    keywords={use,let,LittleEndian,Unit,is,Adj,operation,for,if,else,in,swap,assert,Length,and,del,Qubit,within,apply,Controlled},
    keywordstyle=\color{deepblue},
}

\renewcommand\thesection{\arabic{section}}

\theoremstyle{definition}

\theoremstyle{definition}

\theoremstyle{definition}

\newcommand{\eq}[1]{\hyperref[eq:#1]{Equation~\ref*{eq:#1}}}
\renewcommand{\sec}[1]{\hyperref[sec:#1]{Section~\ref*{sec:#1}}}
\DeclareRobustCommand{\app}[1]{\hyperref[app:#1]{Appendix~\ref*{app:#1}}}
\newcommand{\fig}[1]{\hyperref[fig:#1]{Figure~\ref*{fig:#1}}}
\newcommand{\tbl}[1]{\hyperref[tbl:#1]{Table~\ref*{tbl:#1}}}
\newcommand{\theoremref}[1]{\hyperref[theorem:#1]{Theorem~\ref*{theorem:#1}}}
\newcommand{\definitionref}[1]{\hyperref[definition:#1]{Definition~\ref*{definition:#1}}}

\begin{document}
\title{Quantum Dictionaries without QRAM}

\date{\today}
\author{Craig Gidney}
\email{craig.gidney@gmail.com}
\affiliation{Google Quantum AI, Santa Barbara, California 93117, USA}

\begin{abstract}
This paper presents an efficient gate-level implementation of a quantum dictionary: a data structure that can store a superposition of mappings from keys to values.
The dictionary is stored as a fixed-length list of sorted address-value pairs, where the length of the list is the maximum number of entries that can be put in the dictionary.
An addressed value can be extracted from (or injected into) the dictionary using $C \cdot (V + 2.5A) + O(V + A + C)$ expected Toffoli gates and $O(V + A)$ auxiliary qubits (where $C$ is the maximum capacity, $A$ is the address width, and $V$ is the value width).
\end{abstract}

\maketitle

\section{Introduction}
\label{sec:introduction}

Recently, Buhrman et al proposed a quantum construction for ``compressed memory'' built on top of QRAM \cite{buhrman2022memory}.
A ``compressed memory'' is a superposed mapping from integer addresses to integer values, with a requirement that the number of non-zero values in the memory is small.
In other words, it's the quantum equivalent of a hash table or a binary search tree; a quantum dictionary.
To guarantee reversibility, the dictionary has a fixed maximum capacity $C$ and a convention that an address is mapped to 0 if and only if it's not in the dictionary.

Because I'm personally skeptical that fault tolerant QRAM with cheap query operations will ever exist, I decided to try to create a quantum dictionary that didn't use QRAM, to get a sense of the cost of querying such a dictionary.
This short note explains the implementation I found.

\section{Implementation}

A notable difficulty, when implementing a quantum dictionary, is ensuring that information about the order that operations occurred (or which operations have occurred) is not sneaking into the data structure.
For example, some classical hash tables handle collisions by using linear probing.
But this makes it possible to infer in what order items with colliding hashes were inserted into the dictionary.
This kind of extra information is disastrous in a quantum data structure, because it stops interference effects from happening in the way that was intended by the user.

For my quantum dictionary implementation I decided to use a sorted list of (address, value) pairs, with unused entries indicated by the special pair (MAX\_ADDRESS, 0).
Because the list is sorted, and addresses other than MAX\_ADDRESS can't appear multiple times, there is only one valid representation for a mapping.
This guarantees no extra information can sneak into the list.
The remaining challenge is to operate on this representation without producing intermediate values that can't be uncomputed.

I tackled the dictionary-implementation problem by focusing on one specific dictionary operation: moving an addressed value out of the dictionary (an operation I call ``extraction'').
The extraction operation takes a dictionary, an address register, and an initially-zero'd output register.
If the address isn't in the dictionary, nothing happens.
The output register stays in the $|0\rangle$ state.
If the address is in the dictionary, the extraction operation removes it from the dictionary and moves its value into the output register.

All dictionary operations can be decomposed into extraction and its inverse (an operation I call ``injection'').
For example, to add an offset to an addressed value inside the dictionary, you extract the addressed value from the dictionary, add the offset into the extracted value, then inject the modified value back into the dictionary at the same address.
Beware that, because extraction \emph{guarantees} the address is no longer in the dictionary at the end of the operation, injection \emph{requires} that the address is not already in the dictionary at the start of the operation.
There is no safe way to write to an address in the dictionary without first extracting the value for that address.

I implemented extraction using a strategy that appends an empty (MAX\_ADDRESS, 0) pair to the end of the (address, value) list, and then iterates through the list while pushing any matching entry to the end.
If there was no match, the (MAX\_ADDRESS, 0) pair will still be at the end of list after iterating through the list.
If there was a match, the last item in the list will have an address equal to the query address and a value that isn't 0.
Also, the entries of the dictionary past the match (including the empty pair that was appended at the start) will have been shifted left, restoring all the required invariants.
The trickiest part of implementing this strategy is figuring out how to uncompute the results of the comparisons checking whether the current entry is a match for the query address or not, after having potentially pushed the entry.
It turns out this can be done by noticing that list entries are out of order, which only occurs while a match is being pushed to the end of the list.

Pseudocode implementing the extraction is in \fig{extract}.
Pseudocode using this method to implement other dictionary methods is in \fig{other-methods}.
Actual working Q\# code implementing these methods is attached to the paper as ancillary files.
The implementation can be tested by running ``dotnet test src/project.csproj'' on a machine with Q\# installed.

\begin{figure}
\begin{lstlisting}[style=qsharp]
/// Moves an addressed value out of the dictionary.
/// Performs: swap output, dict[address]
/// Requires: output == 0
/// Ensures: dict[address] == 0
/// Ensures: output == 0 or HasSpace(dict)
operation dict_extract(
        dict: QuantumDict,
        address: LittleEndian,
        output: LittleEndian) : Unit is Adj {
    assert output == 0;
    let a = Length(address!);
    let max_address = (1 <<< a) - 1;

    // Temporarily append an extra empty entry to the dictionary's list.
    use extra_address = LittleEndian(max_address, a);
    let addrs = dict::addrs + [extra_address];
    let vals = dict::vals + [output];

    // Search for the match; pushing it to the end of the list.
    for k in 0..Length(vals)-2 {
        // Check if the current entry matches the target address.
        use eq = Qubit(address == addrs[k]);

        // Push any match towards the end of the list.
        if eq {
            swap vals[k], vals[k+1];
            swap addrs[k], addrs[k+1];
        }

        // Uncompute eq using sort order violation from the swaps.
        del eq = addrs[k] > addrs[k+1];
    }

    // Uncompute the leftover extra address.
    del extra_address = output == 0 ? address | max_address;
}
\end{lstlisting}
\caption{
Q\#-like pseudocode for pulling an addressed value out of a quantum dictionary.
Actual Q\# code is included in the ancillary files attached to this paper (in the ``src/'' directory).
}
\label{fig:extract}
\end{figure}

\begin{figure}
\begin{lstlisting}[style=qsharp]
/// Moves an addressed value into the dictionary.
/// Performs: swap output, dict[address]
/// Requires: dict[address] == 0
/// Requires: HasSpace(dict) or value == 0
/// Ensures: value == 0
operation dict_inject(
        dict: QuantumDict,
        address: LittleEndian,
        value: LittleEndian) : Unit is Adj {
    Adjoint dict_extract(dict, address, value);
}

/// Performs: swap dict[address], value
/// Requires: HasSpace(dict) or ((value != 0) >= (dict[address] != 0))
operation swap_dict_value_for_value(
        dict: QuantumDict,
        address: LittleEndian,
        value: LittleEndian) : Unit is Adj {
    use temp = LittleEndian(0, Length(value!));
    dict_extract(dict, address, temp);
    swap value, temp;
    dict_inject(dict, address, temp);
}

/// Performs: dict[address] += value
/// Requires: HasSpace(dict) or ((value+dict[address]!=0) >= (dict[address]!=0))
operation add_value_into_dict_value(
        value: LittleEndian,
        dict: QuantumDict,
        address: LittleEndian) : Unit is Adj {
    use temp = LittleEndian(0, Length(value!));
    dict_extract(dict, address, temp);
    temp += value;
    dict_inject(dict, address, temp);
}

/// Performs: value += dict[address]
operation add_dict_value_into_value(
        dict: QuantumDict,
        address: LittleEndian,
        value: LittleEndian) : Unit is Adj {
    use temp = LittleEndian(0, Length(value!));
    dict_extract(dict, address, temp);
    value += temp;
    dict_inject(dict, address, temp);
}
\end{lstlisting}
\caption{
Q\#-like pseudocode showing how to decompose various dictionary operations into extraction and its inverse.
Actual Q\# code is included in the ancillary files attached to this paper (in the ``src/'' directory).
}
\label{fig:other-methods}
\end{figure}

My implementation of the extract operation has an expected Toffoli count of $C \cdot (2.5A + V) + O(C + A + V)$ where $C$ is the capacity of the dictionary, $V$ is the number of qubits in a value, and $A$ is the number of qubits in an address.
The cost of $C \cdot (2.5A + V)$ comes from the main loop.
One $A$ comes from computing ``eq'' by checking if the query address is equal to the current pair's address.
One $A$ and one $V$ comes from swapping the current pair for the next pair, controlled by ``eq''.
The remaining $0.5 A$ comes from uncomputing ``eq'', using measurement based uncomputation.
The measurement based uncomputation first measures ``eq'' in the X basis.
Half of the time the result of the measurement is $|+\rangle$ and no further action is needed.
The other half of the time, the result is $|-\rangle$ and states where the current pair's address and the next pair's address are out of order need to have their amplitudes negated.
This can be done using a ripple carry comparator, which has the same cost as a ripple carry adder \cite{gidney2018addition} (in this case: $A + O(1)$ Toffoli gates).

An operation that involves editing a value in the dictionary, like the operation \\``$\text{dict}[\text{address}]\;\text{+=}\;5$'', will have a total cost of $C \cdot (5A + V) + O(C + A + V)$, because it has to perform an extraction and an injection.

The extraction code I've provided has linear depth: $O(C + A + V)$.
It's possible to reduce the depth from linear to logarithmic by using more workspace and more Toffoli gates.
I don't bother doing so in this paper because I expect even the linear depth implementation I've presented to be bottlenecked waiting for magic states \cite{gidney2020blockadder}.
A logarithmic depth implementation that uses more Toffoli gates will actually take longer to execute because it needs to wait for more magic states.

\section{Conclusion}

In this paper I presented an implementation of a quantum dictionary that doesn't use QRAM.
The implementation is reasonably simple (its pseudocode covers half a page), and has a Toffoli count comparable to the gate cost of implementing a quantum circuit that performs a QRAM read.

\section{Acknowledgements}

We thank Hartmut Neven for creating an environment where this work was possible.

\bibliographystyle{plainnat}
\bibliography{refs}

\end{document}